# Angular distribution of thick-target bremsstrahlung produced by electrons with initial energies ranging from 10 to 20-keV incident on Ag

D. Gonzales, B. Cavness, and S. Williams

Department of Physics, Angelo State University, San Angelo, Texas 76909, USA

Experimental results are presented comparing the intensity of the bremsstrahlung produced by electrons with initial energies ranging from 10 to 20-keV incident on a thick Ag target, measured at forward angles in the range 0° to 55°. When the data are corrected for attenuation due to photon absorption within the target, the results indicate that the detected radiation is distributed anisotropically only at photon energies, $k$, that are approximately equal to the initial energy of the incident electrons, $E_o$. The results of our experiments suggest that as $k/E_o \to 0$, the detected radiation becomes essentially isotropic due primarily to the scattering of electrons within the target. A comparison to the theoretical work of Kissel et al. [6] suggests that the angular distribution of bremsstrahlung emitted by electrons incident on thick targets is similar to the angular distribution of bremsstrahlung emitted by electrons incident on free-atom targets only when $k/E_o \approx 1$. The experimental data are also in approximate agreement with the angular distribution predictions of the Monte Carlo program, PENELOPE.

## I. INTRODUCTION

Information concerning the angular distribution of bremsstrahlung is important in such fields as radiation physics, nuclear physics, radiation oncology, and astronomy. While there have been a few experimental studies [1-4] related to the angular distribution of bremsstrahlung emitted by electrons, there is still a need for additional experimental data to compare with theoretical treatments [5, 6] and Monte Carlo simulations [7, 8].

A previous report by our group [4] presented the results of a study of the angular distribution of 10-keV bremsstrahlung produced by 17.5-keV electrons incident on a thick Ag target at forward angles in the range 0°-55°. Corrections were made for several factors including attenuation within the target material. While the results indicated that the bremsstrahlung intensity increased slightly as the angle approached 45° and then decreased slightly thereafter, the increase/decrease was small relative to the uncertainty in our measurements. Thus, our conclusion was that the 10-keV bremsstrahlung resulting from 17.5-keV electrons incident on thick Ag was distributed isotropically. However, the results of experiments performed by Chervenak and Liuzzi involving 30-keV electrons incident on thick Cu suggest that the resultant bremsstrahlung was anisotropic [1]. While Chervenak and Liuzzi made no corrections for attenuation due to absorption in the target, they noted that for geometries where this attenuation was equal some angular dependence was still observed. Furthermore, the theoretical work of Kissel et al. (KQP) [6] suggests that the angular distribution of bremsstrahlung is strongly anisotropic at certain photon-to-electron energy ratios. The purpose of this study is to revisit the subject and to examine the angular distribution of bremsstrahlung for several photon-to-electron energy ratios and incident electron energies.

The theoretical tabulations of KQP [6] include bremsstrahlung energy spectra and shape functions for incident electron energies from 1-keV to 500-keV and target materials with atomic numbers from Z=1 to Z=92. The theoretical doubly-differential bremsstrahlung cross sections and shape functions in the tabulations of KQP [6] are for situations where the target is thin enough that each incident electron only interacts with a single target-atom. The shape function tabulations were based upon "benchmark" theoretical calculations for materials that included Ag which were fitted to the angular distribution as predicted by the Born approximation:

$$S = A / [4\pi (1 - \beta_1 \cos\theta)^m] \sum_{i=0}^{N} B_i P_i (\cos \theta) \qquad (1)$$

where $B_o = 1$, $\beta_1$ is the velocity of the incident electron divided by the speed of light, and $P_i (\cos \theta)$ are Legendre polynomials. A good overall fit was obtained by assigning values of 5 and 4.5 to $N$ and m, respectively. The value of the parameter A was determined through normalization.

The results of experiments involving the angular distribution of bremsstrahlung emitted by electrons incident on free Ar, Kr, and Xe atoms performed by Aydinol et al. [2] support the theoretical work of Tseng et al. [5] and KQP [6].

As mentioned previously, so-called thin-target bremsstrahlung results from each incident electron interacting only with a single target-atom. Thick-target bremsstrahlung (TTB), however, results from each incident electron interacting with multiple target-atoms. Thus, for TTB, the ratio of the energy of the emitted photon to the kinetic energy of the electron just before the photon is emitted is usually different for each bremsstrahlung photon of energy $k$. For example, the 10-keV bremsstrahlung studied in our previous report [4] had been emitted by electrons that had likely already interacted with other target-atoms and lost kinetic energy in the process. For thin-target bremsstrahlung, however, the ratio of the energy of the emitted photon to the kinetic energy of the emitting electron is the same for each bremsstrahlung photon of energy $k$ because each photon is emitted by an electron with the same kinetic energy, $E_o$. Perhaps more importantly, after an initial interaction, the direction of travel for each incident electron is altered when the electron is scattered. Thus, TTB is usually emitted by electrons traveling in a direction that is different than the one they were moving in initially.

Intuitively, it seems that as $k/E_o \rightarrow 1$, the angular distribution of TTB (neglecting any attenuation due to the target material) should approach the theoretical shape functions of KQP. This is due to the fact that a photon of energy $k = E_o$ can only be emitted by an electron during its first interaction with a target-atom (typically, near the surface of the target). As $k/E_o \rightarrow 0$, the angular distribution of the TTB is essentially a superposition of several thin-target shape functions because the electrons emitting the radiation have various kinetic energies that are less than $E_o$, due to previous interactions within the target. If attenuation due to absorption within the target is not considered, it would also seem that this superposition of shape functions, along with the fact that the majority of the electrons emitting radiation have had their initial direction of travel altered, should lead to TTB being essentially isotropic at values of $k$ that are significantly less than $E_o$. Additionally, the KQP tabulations predict a weaker anisotropy as the photon-to-electron energy ratio approaches zero [6].

A FORTRAN-based Monte Carlo program, PENELOPE, designed to simulate bremsstrahlung emission, was developed by Salvat et al. [7] and was partially based upon the KQP tabulations. The shape functions of the KQP tabulations are not, however, incorporated into the PENELOPE code (the way in which PENELOPE predicts the angle of the emission of bremsstrahlung photons has been described by Acosta et al. [9]). The PENELOPE program allows users to specify incident particle type (electrons, positrons, or photons), incident particle energy, target material, and target-thickness. PENELOPE is capable of simulating both thin-target and TTB and the predictions of the code have been shown to be in good agreement with experimental results [9-14].

## II. EXPERIMENTAL PROCEDURE

The data presented in this work were obtained using an experimental setup similar to the one described in our previous report [4]. Electrons were incident on a 2415 µg/cm$^2$ Ag film (positioned at the center of a rotatable stage) and the resultant radiation was detected at forward angles in the range 0° to 55° (i.e. in "transmission mode"). The target's thickness is well beyond the CSDA range for 10-20 keV electrons [15]. X-rays were detected using an LD Didactic Si-PIN detector (except for data related to 17.5-keV electrons, which were taken using a Princeton Gamma-Tech Si(Li) detector).

## III. RESULTS AND DISCUSSION

### A. Angular distribution of the TTB produced by $E_o$=10 keV electrons and comparison to KQP shape function

Figure 1 is a comparison of the angular distributions of bremsstrahlung produced by 10-keV electrons incident on thick Ag for $k/E_o$ values of 0.7, 0.8, 0.9, and 0.95. The plot is of the ratios of the intensities of bremsstrahlung at an angle θ to the intensities at 0° at angles ranging from 0° to 55°. Attenuation due to absorption within the target was corrected for by applying an exponential term:

$$f(E_o, k, \theta) = e^{(\mu_k x/\cos\theta)} \qquad (2)$$

where $\mu_k$ is the attenuation coefficient of a photon with energy $k$ for Ag (taken from the NIST XCOM database[16]) and $x/\cos\theta$ is the distance that the photon travels through the target material. This distance is determined through an approximation that assumes that an individual electron's energy loss is linear as it travels through the target and that at the CSDA range the average electron-energy is 0-keV. The "scaled" bremsstrahlung energy spectrum tabulations of Pratt et al. [17] (based on a single-interaction model) suggest that for the range of electron energies related to our experimental data, the likelihoods of an electron emitting a photon of any energy ranging from its kinetic energy to 0-keV are approximately equal. Thus, the *average* point from which a photon of energy $k$ originates in the target is approximately the mid-point between the target surface and the point at which the CSDA range predicts the average electron energy to be equal to $k$. The error in the experimental data shown in Figures 1 and 2 is estimated to be less than 3% for all data points. The error is independent of detector efficiency and solid angle and is estimated by combining statistical error and the estimated error in energy calibration in quadrature.

As one might expect, the plot indicates that the anisotropy of the detected TTB is much stronger for $k/E_o$ values near unity and that the ratios $I_\theta/I_{0°}$ for $k/E_o$ values of 0.7 and 0.8 are only about 10% higher at 55° than at 0°. The general trend seen in Figure 1 is in agreement with the theoretical work of KQP [6], which predicts that the intensity of the bremsstrahlung should increase as the angle of emission is increased, up until about 70°, at which point the intensity is at a maximum. However, as indicated in Figure 2, the anisotropy of our TTB experimental data is not nearly as strong as predicted by the thin-target KQP tabulations for $k/E_o$ = 0.95. The KQP data predicts an $I_\theta/I_{0°}$ ratio that is almost 3 times greater at 55° than it is at 0°, while the ratio $I_\theta/I_{0°}$ calculated using the TTB experimental data is slightly less than 1.5 times greater at 55° than it is at 0°. This suggests that even for values of $k/E_o$ that are slightly less than unity, electron scattering must be taken into consideration when modeling TTB.

**B. Angular distribution of the TTB produced by $E_o$=17.5 keV electrons and comparison to the predictions of PENELOPE**

Figure 3 is a comparison of the angular distributions of bremsstrahlung produced by 17.5-keV electrons incident on thick Ag for $k/E_o$ values of 0.5, 0.7, 0.9, and 0.95. As in Figure 1, the plot is of the ratios of the intensities of bremsstrahlung at an angle θ to the intensities at 0° as a function of angle of emission, and attenuation due to absorption within the target has been corrected using Equation 2. The graph indicates that the anisotropy of the detected TTB is much stronger for $k/E_o$ values of 0.95 and 0.9 than it is for $k/E_o$ values of 0.7 and 0.5. The $I_\theta/I_{0°}$ ratio for $k/E_o$ = 0.95 is slightly more than 21% higher at 55° than it is at 0°. Once again, the anisotropy observed in our TTB experiments is not as strong as predicted for situations involving electrons incident on free-atom targets by the KQP tabulations; however, the general trend seen in Figure 3 is in agreement with the KQP tabulations. These results are consistent with those of Chervenak and Liuzzi's experiments involving 30-keV electrons incident on thick Cu [1]. They attribute the anisotropy observed in their experiments to thin-target bremsstrahlung production and their results indicate a difference in intensities of no more than about 25%. In our data, the same relatively weak anisotropy (in comparison to the predictions of KQP) is observed even as $k/E_o \to 1$. The error in the experimental data shown in Figure 3 is estimated to be no more than 2% for all data points. The error is estimated by combining statistical error and the estimated error in energy calibration in quadrature.

Figures 4 and 5 are comparisons of the absolute bremsstrahlung probability densities for 17.5-keV electrons incident on thick Ag as predicted by the Monte Carlo code, PENELOPE, and as measured through experiment as a function of the angle of detection for $k/E_o$ values of 0.95 and 0.5, respectively. The absolute probability densities were calculated using the following expression:

$$p(k) = N_p / (N_e \, \Delta k \, \Delta\Omega \, \varepsilon(k)) \qquad (3)$$

where $N_p$ is the number of bremsstrahlung photons detected, $N_e$ is the number of radiating electrons incident on the target, $\Delta k$ is the energy range in which the photons were detected (eV), $\Delta\Omega$ is the solid angle element (sr), and $\varepsilon(k)$ is the efficiency of the detector at energy $k$. Also included is a term, R, to correct for average electron-backscattering. This term was calculated according to the empirical expression used by Ambrose et al. [3]:

$$R(k, E_o) = (1 - \eta) / [1 - \eta \, (k/E_o)^2] \qquad (4)$$

where η is the electron backscattering coefficient (dependent upon $E_o$) for Ag, determined using the expression introduced by August and Wernisch [18]. Thus, $N_e$ was calculated using the expression:

$$N_e = [I \, t \, R(k, E_o)] / e \qquad (5)$$

where I is the electron beam current, t is the time that the electron beam was incident on the target, and *e* is the charge of an electron.

The errors in the experimental data shown in Figures 4 and 5 are estimated to be no more than approximately 17.5% for all data points. The errors have been determined by combining the statistical error and the uncertainties in incident charge, solid angle, detector efficiency, and energy calibration in quadrature. The uncertainties related to the data generated by the PENELOPE code represent three standard deviations and were generated by the program itself.

The probability density predictions of the PENELOPE code take attenuation due to absorption within the target material into consideration, thus the data generated by the program had to be "corrected" using Equation 2 (just as the experimental data was).

The experimental data in Figure 4 shows the same ~21% difference in the bremsstrahlung intensities at angles of 0° and 55° as shown in Figure 3 (however, the scaling makes the increase more subtle). The statistical nature of the data generated by PENELOPE makes any trend in the data difficult to discern. However, the code seems to correctly predict that even for a $k/E_o$ value of 0.95, the anisotropy of the emitted radiation is much weaker than as predicted for situations involving electrons incident on free-atom targets by the KQP tabulations. In general, the predictions of PENELOPE seem to be in good agreement with the experimental data at forward angles in the range 0°-55°.

The experimental data shown in Figure 5 suggests that as $k/E_o \rightarrow 0$, TTB is essentially distributed isotropically. The 8.75-keV bremsstrahlung produced by 17.5-keV electrons incident on thick Ag is emitted by electrons that have likely interacted with multiple target atoms prior to emission, and have thus essentially had the directions that they are traveling in randomized. The probability densities shown in Figure 5 differ by less than 3.5%. Figure 5 also shows that the PENELOPE program is capable of accurately simulating the isotropic TTB distribution for values of $k/E_o$ that are significantly less than unity.

**C. Angular distribution of the TTB produced by $E_o$=20 keV electrons and comparison to the predictions of PENELOPE**

Figure 6 shows a comparison of the absolute bremsstrahlung probability densities for 20-keV electrons incident on thick Ag as predicted by PENELOPE and as measured through experiment as a function of the angle of detection for a $k/E_o$ value of 0.95. The error in the experimental data shown in Figure 6 is estimated to be no more than approximately 12% for all data points. The error has been determined by combining the statistical error and the uncertainties in incident charge, solid angle, detector efficiency, and energy calibration in quadrature. Both data sets have been corrected for attenuation due to absorption within the target.

While the data related to experiments involving 10 and 17.5-keV electrons suggest that for photon-to-incident electron energy ratios of 0.95 the angular distribution of the TTB is anisotropic and similar to that of thin-target bremsstrahlung, the data in Figure 6 does not. The probability densities shown in Figure 6 differ by no more than about 8%, which is well within the uncertainties of our measurements. The reason for the observed isotropy may be that the 19-keV photons were, on average, emitted from a region in the target that is further from the surface than where the $k/E_o$=0.95 photons were emitted, on average, in the 10 and 17.5-keV experiments. Thus, fewer of the electrons that emitted the radiation were traveling in their initial direction and little to no anisotropy due to thin-target bremsstrahlung is observed.

The predictions of the PENELOPE code are in agreement with the experimental data shown in Figure 6. As in Figures 4 and 5, the statistical nature of the data generated by PENELOPE makes it difficult to discern any trends in the code's predictions related to the angular distribution of the 19-keV bremsstrahlung. However, PENELOPE seems to correctly predict that the angular distribution of the 19-keV TTB produced by 20-keV electrons incident on thick Ag is essentially isotropic.

**IV. CONCLUSION**

Results have been presented comparing the intensities of the bremsstrahlung produced by electrons with energies ranging from 10 to 20-keV incident on a thick Ag target, measured at forward angles in the range 0° to 55°. When the data were corrected for attenuation due to photon absorption within the target, the results indicated that electron scattering within the target results in the apparent isotropic distribution of TTB for photon energies $k$ that are not approximately equal to the initial energy of the incident electrons, $E_o$. However, for photon energies $k$ that are approximately equal to $E_o$, the intensity of the bremsstrahlung increases as the angle of detection is increased at forward angles in the range 0° to 55°. Our experimental results also suggest that as the energy of the incident electrons is increased, the $k/E_o$ ratio must be closer to unity in order for the observed bremsstrahlung to be distributed anisotropically.

A comparison of data obtained from experiments involving 10-keV electrons incident on thick Ag to the theoretical thin-target shape functions of KQP [6] indicates that for a $k/E_o$ value of 0.95, the general trend in the angular distributions of the radiation are similar. However, the anisotropy observed was much weaker than as predicted by the KQP tabulations for free-atom targets. This is likely due to electron scattering in the target near the surface. The KQP tabulations include shape functions for $k/E_o=1$ (which were estimated from a linear extrapolation of the data for lower $k/E_o$). The authors of the KQP tabulations mention that the $k/E_o=1$ shape functions are meant to be apply to bremsstrahlung photons that are "50 eV or so from the exact tip of the spectrum" [6]. We were unable to compare our experimental data to the $k/E_o=1$ shape functions due to the high uncertainties that would result from poor statistics and detector pile-up. However, based on our experimental data, it seems likely that the angular distribution of TTB would be in approximate agreement with the thin-target KQP tabulations at $k/E_o=1$.

Comparisons of the absolute bremsstrahlung probability densities for 17.5 and 20-keV electrons incident on thick Ag as predicted by the Monte Carlo code, PENELOPE, and as measured through experiment as a function of the angle of detection suggest that PENELOPE is relatively capable of making accurate predictions. However, the statistical nature of the data generated by PENELOPE made it difficult to discern any trends related to the angular distributions of the bremsstrahlung.

In the future, additional theoretical work related to the angular distribution of TTB needs to be performed (as well as additional experimental work with which to make comparisons). Also, it would be interesting to perform experiments involving free-atom targets (such as those performed by Aydinol et al. [2]) in order to study the angular distribution of polarizational bremsstrahlung [19], which has not been observed in solid-target experiments [12].

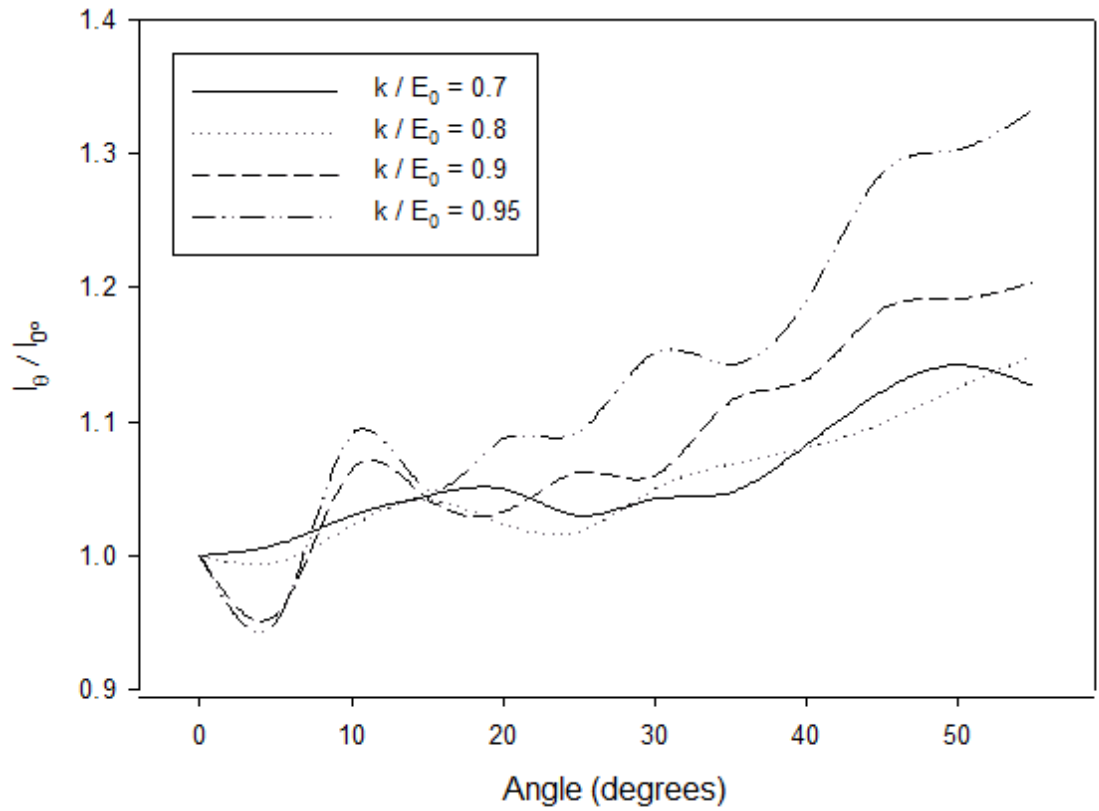

FIG. 1. Comparison of the angular distributions of bremsstrahlung produced by 10-keV electrons incident on thick Ag for $k/E_o$ values of 0.7, 0.8, 0.9, and 0.95.

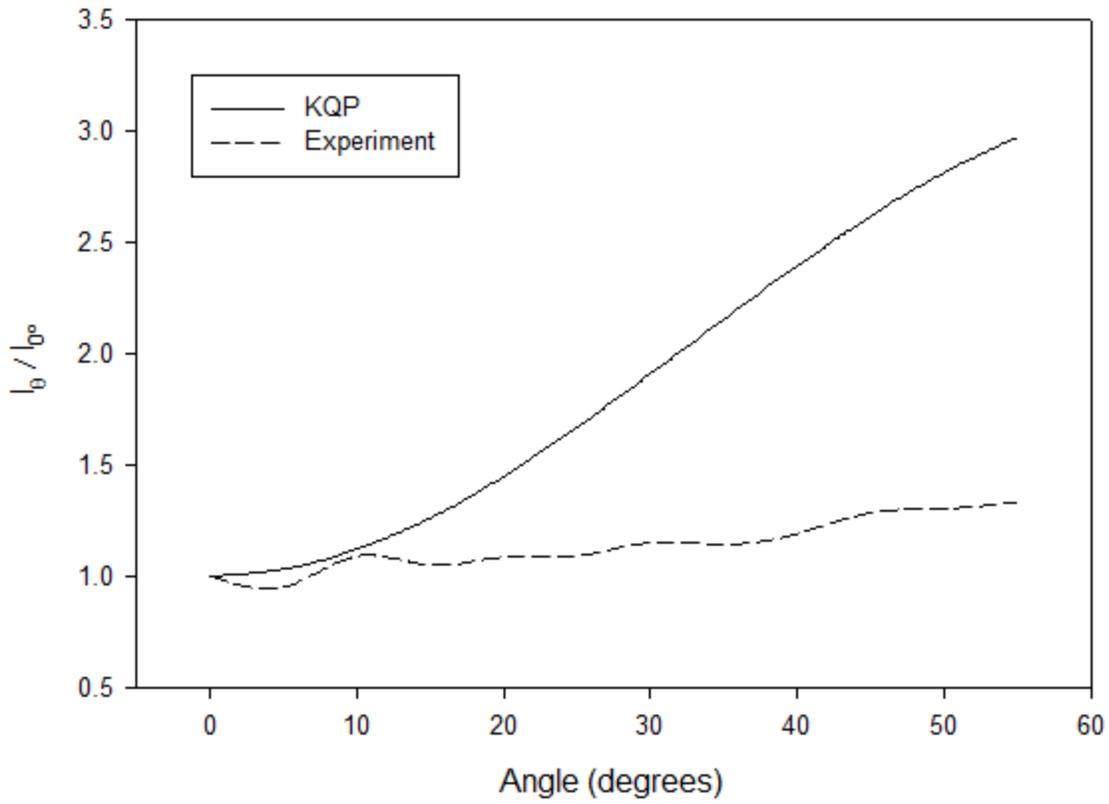

FIG. 2. Comparison of data obtained from experiments involving 10-keV electrons incident on thick Ag to the theoretical thin-target shape functions of KQP for $k/E_o$=0.95.

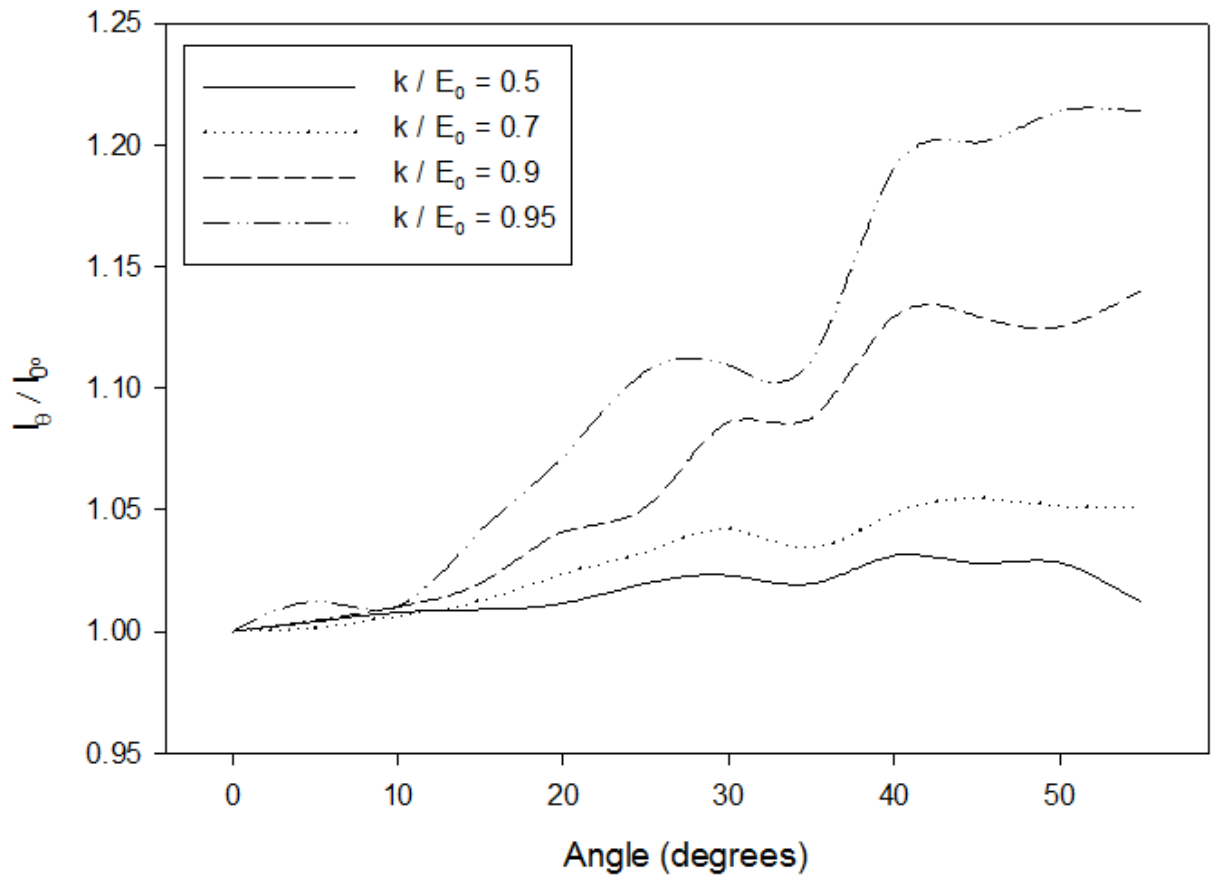

FIG. 3. Comparison of the angular distributions of bremsstrahlung produced by 17.5-keV electrons incident on thick Ag for $k/E_o$ values of 0.5, 0.7, 0.9, and 0.95.

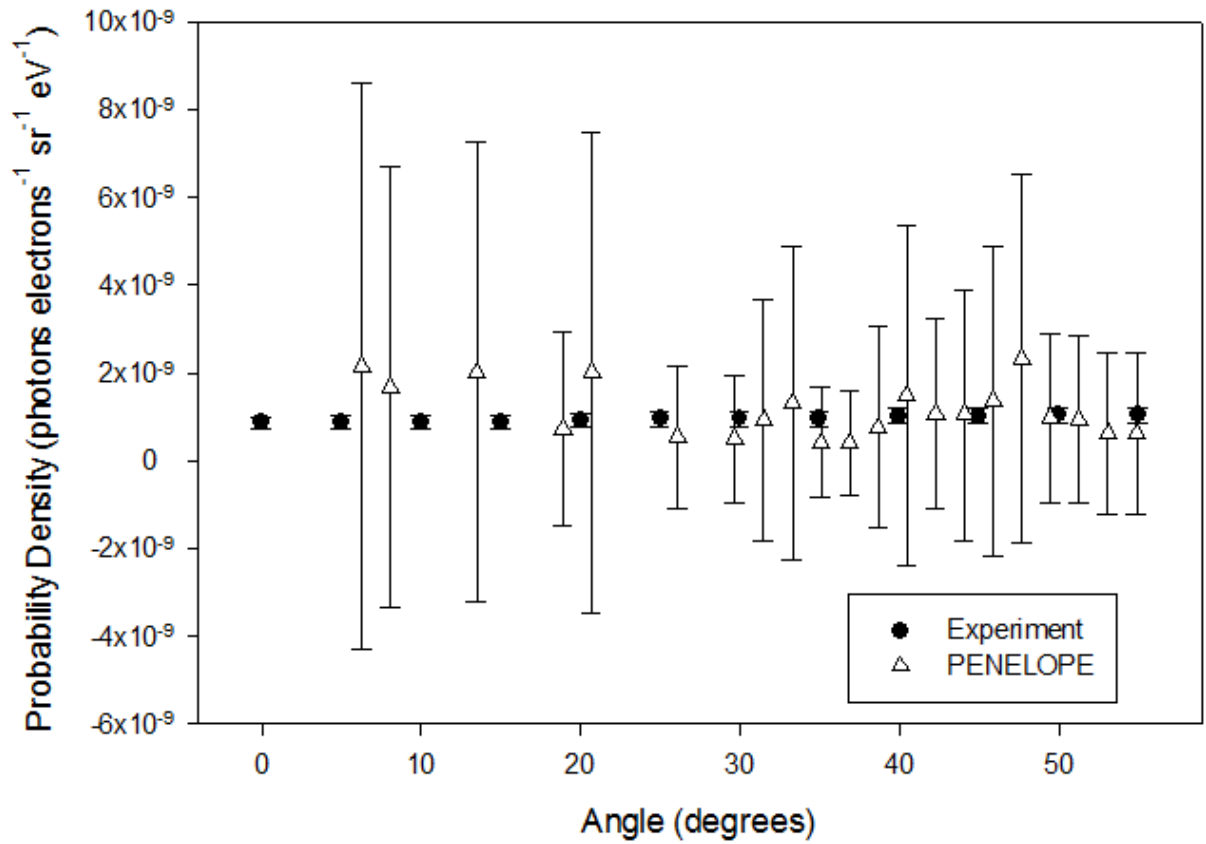

FIG. 4. Comparison of the absolute bremsstrahlung probability densities for 17.5-keV electrons incident on thick Ag as predicted by the Monte Carlo code PENELOPE and as measured through experiment as a function of the angle of bremsstrahlung emission for $k/E_o=0.95$.

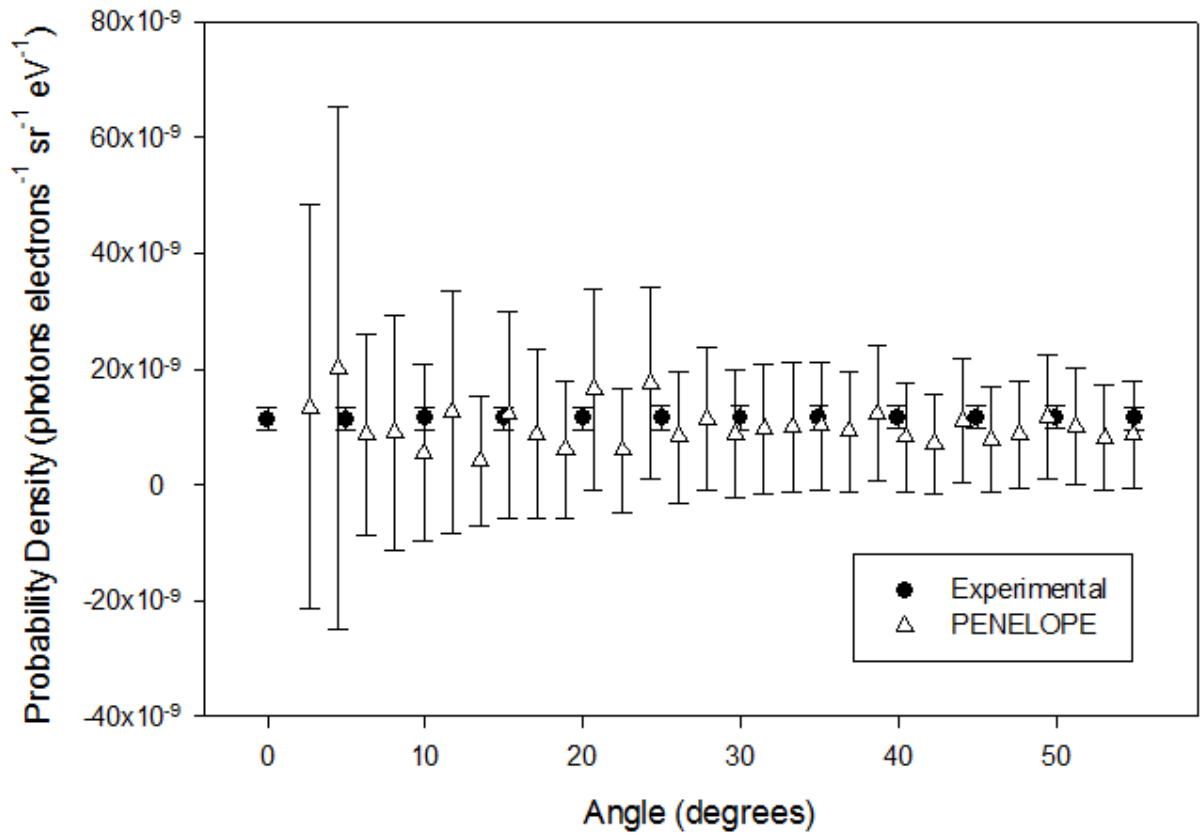

FIG. 5. Comparison of the absolute bremsstrahlung probability densities for 17.5-keV electrons incident on thick Ag as predicted by the Monte Carlo code PENELOPE and as measured through experiment as a function of the angle of bremsstrahlung emission for $k/E_o=0.5$.

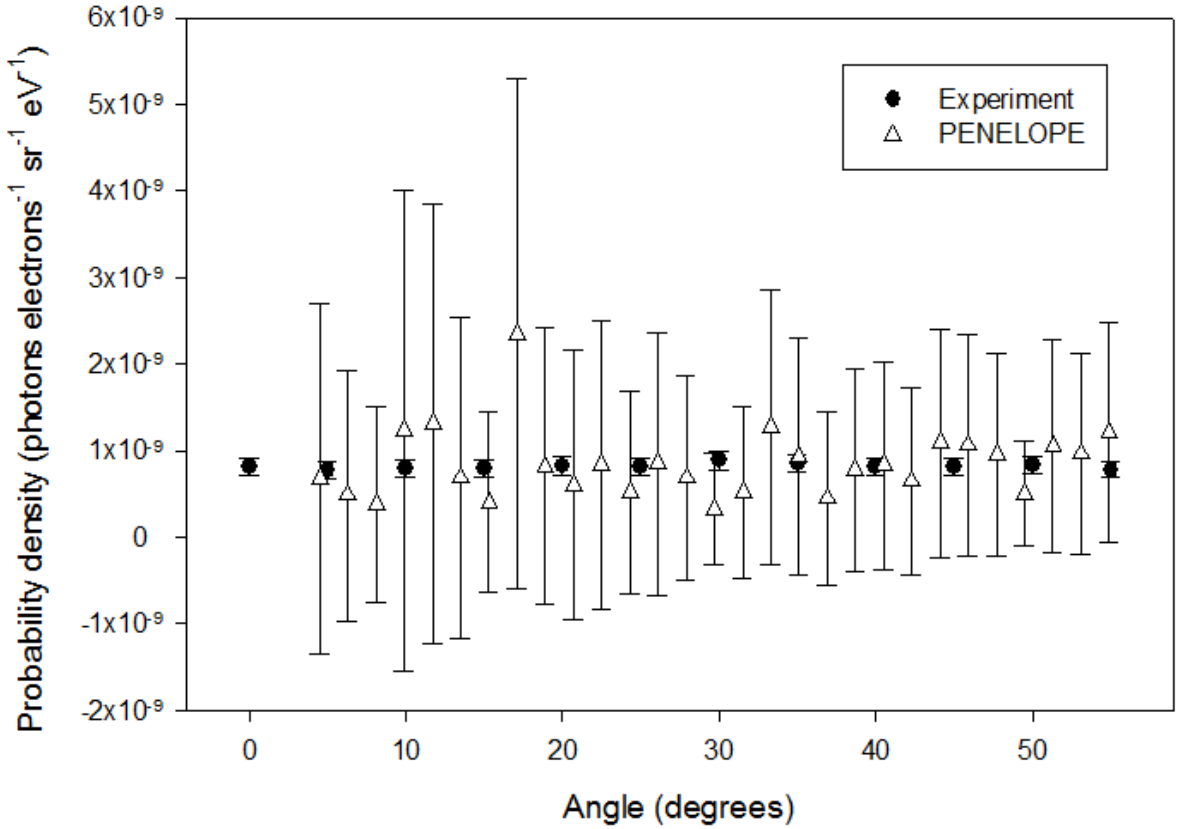

FIG. 6. Comparison of the absolute bremsstrahlung probability densities for 20-keV electrons incident on thick Ag as predicted by the Monte Carlo code PENELOPE and as measured through experiment as a function of the angle of bremsstrahlung emission for $k/E_o=0.95$.